\pdfoutput=1

\documentclass[
superscriptaddress,
twocolumn,
amsmath,amssymb,
aps,
prb,
floatfix,
showkeys,
]{revtex4-2}

\usepackage{graphicx}
\usepackage{dcolumn}
\usepackage{bm}
\usepackage{amsmath}
\usepackage{amssymb} 
\usepackage{multirow}
\usepackage{xcolor}
\usepackage{booktabs}
\usepackage{tabularx}
\usepackage[english]{babel}
\usepackage[utf8]{inputenc}
\usepackage{enumerate}
\usepackage{newunicodechar}
\newunicodechar{ﬁ}{fi}
\newunicodechar{ﬀ}{ff}
\usepackage[colorlinks, citecolor=blue, linkcolor=blue]{hyperref}

\graphicspath{{figs/}}

\renewcommand{\selectlanguage}[1]{}

\begin{document}
\title{Strain-driven domain wall network with chiral junctions in an antiferromagnet}

\author{Vishesh Saxena}
\affiliation{Institute of Nanostructure and Solid State Physics, University of Hamburg, Jungiusstrasse 11, 20355 Hamburg, Germany}

\author{Mara Gutzeit}
\affiliation{Institute of Theoretical Physics and Astrophysics, University of Kiel, Leibnizstrasse 15, 24098 Kiel, Germany}

\author{Arturo Rodríguez-Sota}
\affiliation{Institute of Nanostructure and Solid State Physics, University of Hamburg, Jungiusstrasse 11, 20355 Hamburg, Germany}

\author{Soumyajyoti Haldar}
\affiliation{Institute of Theoretical Physics and Astrophysics, University of Kiel, Leibnizstrasse 15, 24098 Kiel, Germany}

\author{Felix Zahner}
\affiliation{Institute of Nanostructure and Solid State Physics, University of Hamburg, Jungiusstrasse 11, 20355 Hamburg, Germany}

\author{Roland Wiesendanger}
\affiliation{Institute of Nanostructure and Solid State Physics, University of Hamburg, Jungiusstrasse 11, 20355 Hamburg, Germany}

\author{Andr\'e Kubetzka}
\affiliation{Institute of Nanostructure and Solid State Physics, University of Hamburg, Jungiusstrasse 11, 20355 Hamburg, Germany}

\author{Stefan Heinze}
\affiliation{Institute of Theoretical Physics and Astrophysics, University of Kiel, Leibnizstrasse 15, 24098 Kiel, Germany}
\affiliation{Kiel Nano, Surface, and Interface Science (KiNSIS), University of Kiel, Germany}

\author{Kirsten von Bergmann}\email[]{kirsten.von.bergmann@physik.uni-hamburg.de}
\affiliation{Institute of Nanostructure and Solid State Physics, University of Hamburg, Jungiusstrasse 11, 20355 Hamburg, Germany}

\date{\today}

\begin{abstract}
Materials with antiferromagnetic order have recently emerged as promising candidates in spintronics based on their beneficial characteristics such as vanishing stray fields and ultra-fast dynamics. At the same time more complex localized non-coplanar magnetic states as for instance skyrmions are in the focus of applications due to their intriguing
properties such as the topological Hall effect. Recently a conceptual shift has occurred to envision the use of such magnetic defects not only in one-dimensional race
track devices but to exploit their unique properties in two-dimensional networks. Here we use local strain in a collinear antiferromagnet to induce non-coplanar domain wall junctions, which connect in a very specific way to form extended networks. We combine spin-polarized scanning tunneling microscopy with density functional theory to characterize the different building blocks of the network, and unravel the origin of the handedness of triple-junctions and the size of the arising topological orbital moments.

\end{abstract}

\maketitle

Antiferromagnets have moved into the focus of application-related research due to their favorable properties for instance regarding ultrafast spin dynamics, lack of stray fields, and abundance in various material classes~\cite{JungwirthNN2016,BaltzRMP2018,Bonbien_2022}. Next to collinear antiferromagnets also crystals with non-collinear magnetic order~\cite{TsaiN2020,HigoN2022} 
or non-coplanar order~\cite{TakagiNN2023,ParkNC2023}
have attracted significant attention in view of future antiferromagnetic spintronics devices. At the same time the absence of a net magnetization and the local compensation of magnetic moments on the atomic scale make experimental investigations difficult~\cite{Heinze2000,CheongQM2020}. In addition to the direction of the Néel vector and the size of domains, 
the properties of antiferromagnetic domain walls are relevant for spintronics applications. The experimental characterization of the internal spin structure of antiferromagnetic domain walls or other magnetic defects is challenging and has been realized mostly by spin-polarized scanning tunneling microscopy (SP-STM)~\cite{BodeNM2006,SpethmannNC2021,LeeAM2024} and scanning NV center microscopy techniques~\cite{WoernlePRB2021,HedrichNP2021,TanNM2024}. 

Depending on the symmetry of the system magnetic materials can have several symmetry-equivalent domains, which can result in domain wall junctions where more than two domains meet. Such junctions have been characterized in systems with lamellar stripe domains~\cite{SchoenherrNP2018,CortesPRB2019,RepickyS2021,BrueningPRB2022,FincoPRL2022} and 
in disordered antiferromagnetic domain wall networks~\cite{LeeAM2024}. Often such localized junctions show non-coplanar spin textures, which exhibit intriguing properties such as their response to spin or charge currents, which can manifest in motion of particle-like states~\cite{fert2013skyrmions}, formation of topological orbital  
moments~\cite{Grytsiuk2020}, and contributions to topological Hall signals~\cite{NeubauerPRL2009,TakagiNN2023,ParkNC2023}. Such localized magnetic objects, extended defects, or junctions can form disordered lattices or networks, and in the form of magnetic skyrmions they have been investigated in the context of novel computing schemes such as neuromorphic or reservoir computing~\cite{PrychyPRA2018,PinnaPRA2000,GrollierNE2020,EverschorNRP2024}.

The position of such localized junctions was found to be associated with a lattice distortion due to a curved sample surface~\cite{RepickyS2021} and strain has also been deliberately introduced to orient stripe domains on the several-micrometer scale~\cite{ZhaoAMa2024}. Also in antiferromagnets a significant magneto-elastic energy is present, which manifest for instance in structural distortions below the Néel temperature~\cite{WarmuthQM2018}, a shape-induced anisotropy of the Néel vector orientation~\cite{MeerPRB2022}, or a preferred domain structure~\cite{GomonayJPCM2002,ConsoloRM2023}. Similarly, the properties of non-collinear antiferromagnets were found to be highly sensitive to strain~\cite{JohnsonAX2023}.

Here we locally introduce strain into a film of a collinear antiferromagnet to generate a network of non-collinear antiferromagnetic domain walls and non-coplanar junctions. The network and its constituents are investigated with SP-STM and density functional theory (DFT). We find a magnetism-induced lateral shift of the atoms on the order of $40\%$ of the atomic distances, leading to chiral strain fields at triple-junctions, where three domains meet in one point. The network is generated by the formation of hexa-junctions at positions of local strain and all connection points of domain walls constitute areas with non-coplanar magnetic order and exhibit 
topological orbital moments.

\section*{Building Blocks of the Domain Wall Network}

\begin{figure*}[htb]
	\centering
    \includegraphics[width=0.85\textwidth]{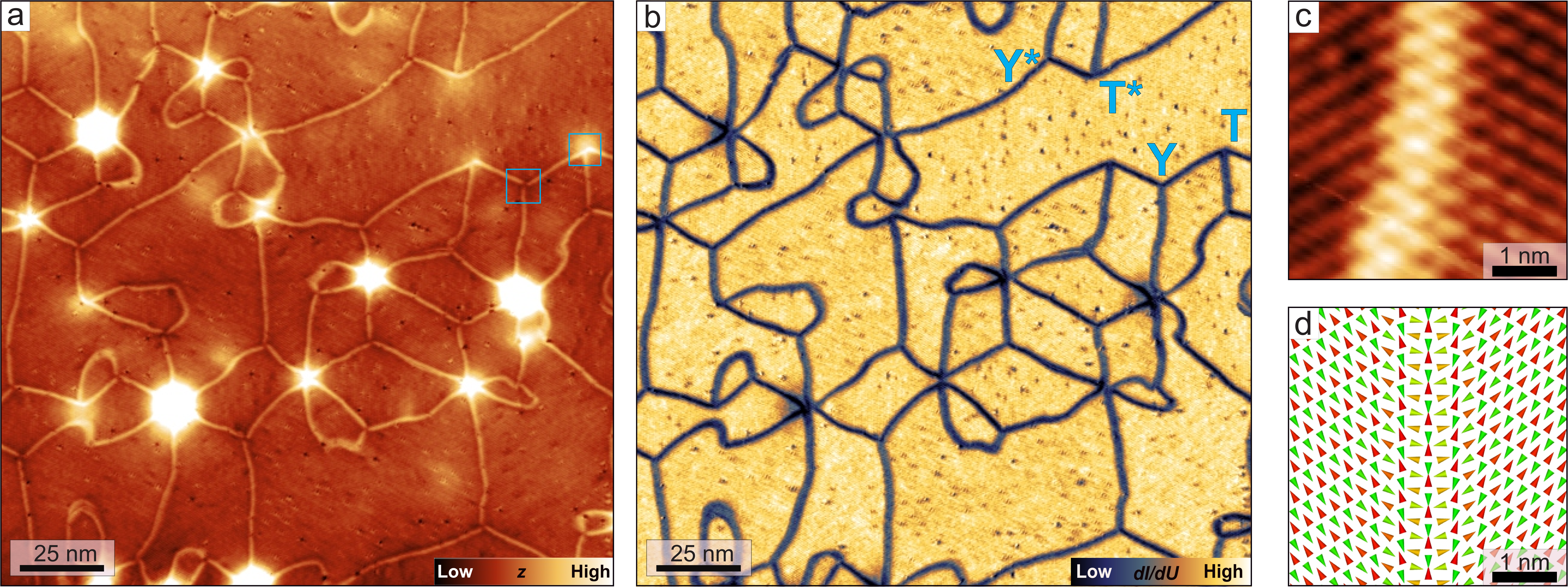}
	\caption{\textbf{Strain-induced domain wall network in a collinear antiferromagnet.} 
    \textbf{a,}~Constant-current STM image of a Mn double-layer film on Ir(111); the bright spots correspond to Ar bubbles below the Ir surface; bright lines indicate domain walls between orientational antiferromagnetic domains ($\Delta z=60$~pm); cyan boxes indicate sample areas which are shown again in Figs.~\ref{fig:walls}c,d.
    \textbf{b,}~d$I$/d$U$~map acquired simultaneously with (a), labels refer to the types of triple-junction.
    \textbf{c,}~SP-STM constant-current image of a domain wall between two orientational antiferromagnetic domains ($\Delta z=15$~pm).
    \textbf{d,}~Spin model of a superposition wall between orientational antiferromagnetic domains (see methods).
    (Measurement parameters: a,b: $U = +10$\,mV, $I = 5$\,nA; c: $U = +10$\,mV, $I = 3$\,nA; all: $T=4$\,K). 
    }
	\label{fig:network}
\end{figure*}

We study an ultra-thin film system that hosts a row-wise antiferromagnetic (RW-AFM) state on a hexagonal lattice, namely a two atomic layer thick Mn film on the (111) surface of an Ir crystal (see methods). In this system the RW-AFM state has three symmetry-equivalent orientational domains. An overview STM constant-current image is presented in Fig.~\ref{fig:network}a. The bright lines that form a network are domain walls separating adjacent orientational domains of the RW-AFM state. The bright spots on the order of $5-10$\,nm in diameter and $15-130$\,pm in height are Ar bubbles located below the surface~\cite{GsellS1998} (see also Suppl. Fig.~1). The domain wall network can also be seen in the simultaneously acquired map of differential conductance (d$I$/d$U$) in Fig.~\ref{fig:network}b. Here the Ar bubbles are nearly invisible and the domain walls appear as dark lines.

A closer view of a domain wall is presented in Fig.~\ref{fig:network}c: for this SP-STM constant-current image a magnetic tip was used, and the resulting tunnel current scales with the projection of tip and sample magnetization (see methods). The RW-AFM state is identified by its characteristic appearance of alternating bright and dark atomic rows~\cite{Wortmann2001,Spethmann2020}, see also illustration of the magnetic state in Fig.~\ref{fig:network}d. The lines visible in the two orientational domains of Fig.~\ref{fig:network}c,d enclose an angle of $120^{\circ}$, and we find that all of the straight walls in the overview image (Fig.~\ref{fig:network}a,b) are of this type. In the center of this domain wall a hexagonal pattern appears and we conclude that they are superposition domain walls as previously observed for the RW-AFM state in the Mn monolayer on Re(0001)~\cite{SpethmannNC2021}. These walls represent a continuous transition between adjacent orientational domains via a superposition state that is characterized by several $90^{\circ}$ angles between neighboring moments, see spin dynamics simulation of such a wall in Fig.~\ref{fig:network}d with generic model parameters (see methods).

\begin{figure}[htb]
	\centering
	\includegraphics[width=0.95\columnwidth]{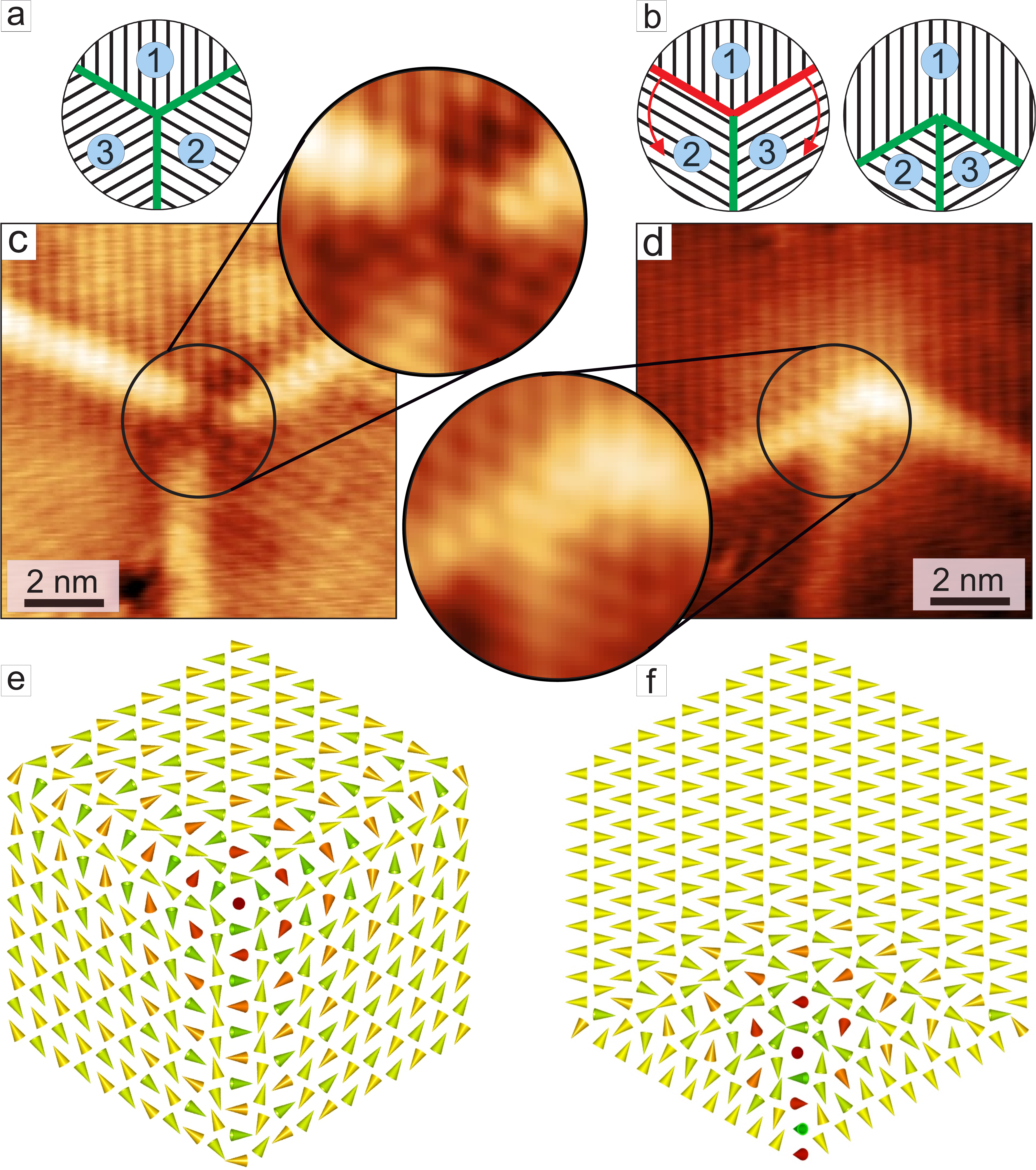}
	\caption{\textbf{Domain wall triple-junctions.} 
    \textbf{a,b,}~Schematics of the two different types of triple junctions, i.e.\ Y and T, respectively; numbers refer to the three possible orientational antiferromagnetic domains, green lines indicate the favorable $120^\circ$ domain walls.  
    \textbf{c,d,}~SP-STM constant-current images of a Y- and a T-junction, respectively ($\Delta z=36$~pm and $\Delta z=25$~pm), see boxes in Fig.~\ref{fig:network}a for their positions within the network. The insets show a magnified view of the central area of the junction exhibiting the hexagonal magnetic pattern characteristic for the 3Q state; for better visibility of the magnetic pattern here high frequency noise has been removed by a lowpass filter with a cut-off frequency corresponding to a wavelength of $0.4$~nm. ($U = +10$\,mV, $I = 5$\,nA, $T=4$\,K).
    \textbf{e,f,}~Spin models of the two types of junctions (see methods). 
    }
	\label{fig:walls}
\end{figure}

The network consists of different connections of domain walls, and two different types of triple domain wall junctions are found as indicated in Fig.~\ref{fig:network}b: either the triple-junction has threefold rotational symmetry, and we will refer to them as Y/Y{$^*$}-junctions, or the threefold symmetry is broken, and we call them T/T{$^*$}-junctions from now on. These types of junctions are reminiscent of some of those previously observed in lamellar stripe domains~\cite{CortesPRB2019,SchoenherrNP2018,FincoPRL2022,RepickyS2021}, and we can rationalize their structure by the sketches in Fig.~\ref{fig:walls}a,b: when all three possible orientational domains meet in one point they can have two different configurations, as indicated by the order of the numbers. With the constraint that all straight walls are $120^{\circ}$ walls (indicated by green lines) we see that only one of the configurations can have a symmetric Y-junction, while the other configuration has to adopt its wall path to form a more asymmetric T-shape. 

Closer-view SP-STM constant-current images of a Y- and a T-junction are displayed in Fig.~\ref{fig:walls}c,d and in both configurations the characteristic $120^{\circ}$ domain walls are seen. We find that, depending on the specific magnetization direction on the employed tip, not only the walls show a hexagonal magnetic pattern but also the center of the junction (see also Suppl.Fig.~2). We would like to note that all triple-junctions in the network displayed in Fig.~\ref{fig:network} have these characteristic patterns. The corresponding sketches in Fig.~\ref{fig:walls}e,f are the result of spin dynamics simulations and show that at the center of the junctions a 3Q state emerges, which is the superposition state of all three adjacent antiferromagnetic domains. This 3Q state, or triple-Q state, is a non-coplanar spin state characterized by tetrahedron angles between all nearest neighbors and has four atoms in the magnetic unit cell~\cite{Kurz2001,Spethmann2020,TakagiNN2023,ParkNC2023,Nickel2023}.

\section*{First-principles calculations of magnetic and structural properties}

\begin{figure}[htb]
	\centering
	\includegraphics[width=1.0\columnwidth]{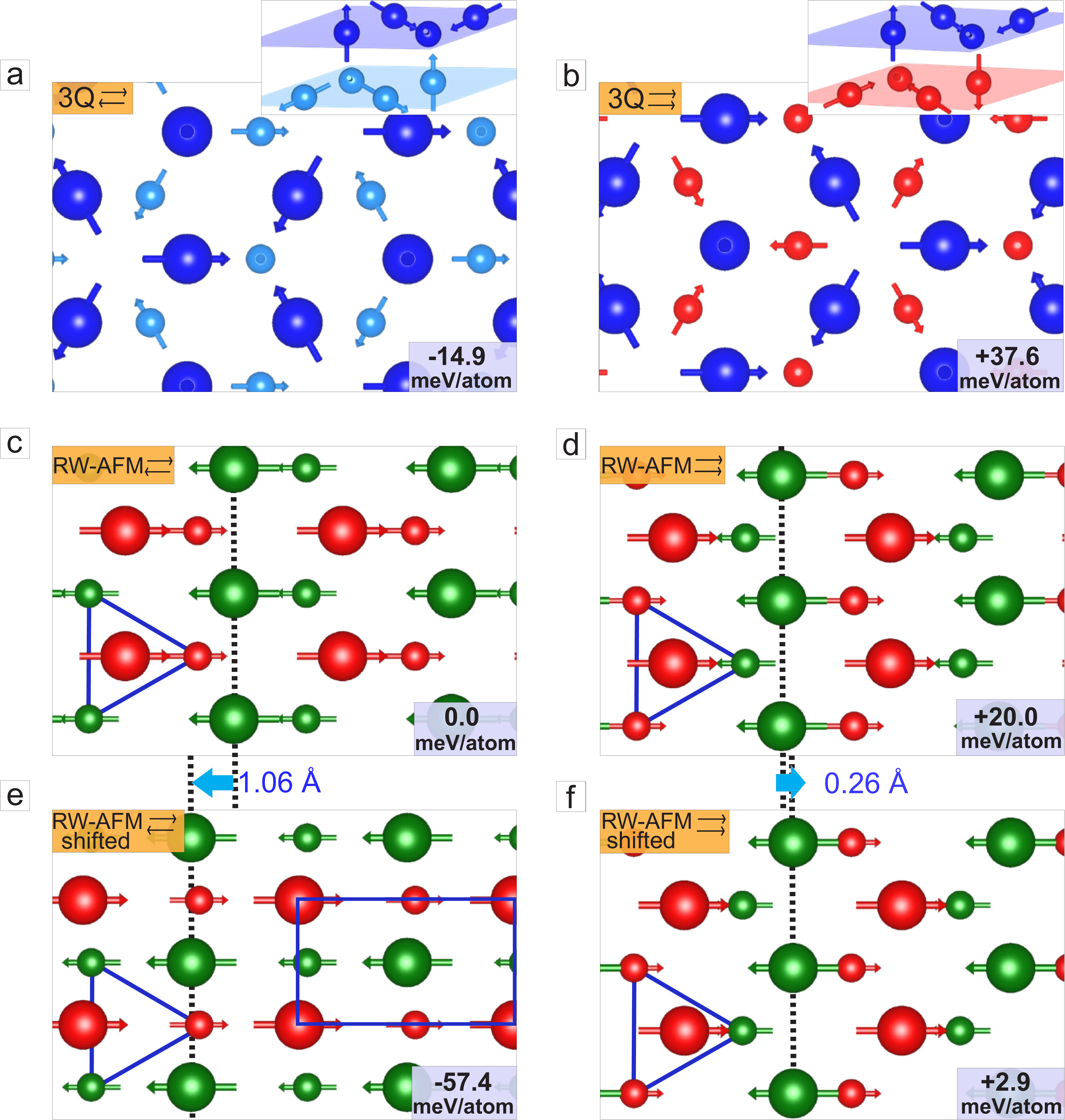}
	\caption{\textbf{Competing spin configurations of the Mn double-layer on Ir(111).} 
    \textbf{a,b,}~Top view of the 3Q state for a net
    AFM ($\rightleftarrows$) and FM ($\rightrightarrows$) coupling between the moments of the Mn layers. The top layer atoms are represented by larger spheres, see also insets for perspective views. 
    \textbf{c,d,}~Top view of the corresponding RW-AFM states; each of them represents one orientation of the 1Q states that lead to the 3Q superposition states in (a,b); blue triangles indicate the triangular lattice of the lower Mn layer.
    \textbf{e,f}~Laterally relaxed RW-AFM states, where the displacement of the Mn atoms in the top layer relative to the hollow sites is indicated by cyan arrows. The total energies of each magnetic structure are given with respect to the RW-AFM$\rightleftarrows$ state.
    } 
	\label{fig:RWAFM_shift}
\end{figure}

To obtain deeper insight into the magnetic properties of this system we employed DFT (see methods). We performed total energy calculations for different magnetic states of the Mn double-layer on Ir(111). Three different spin configurations have been considered within each atomic layer, namely the ferromagnetic (FM), the RW-AFM, and the (non-coplanar) 3Q state, and for each of these states both an effective ferromagnetic ($\rightrightarrows$) and antiferromagnetic ($\rightleftarrows$) alignment between the layers was taken into account, leading to six spin configurations. The states with FM order in each layer have by far the highest total energy as expected for Mn (for values see Table S2 in the Supplementary Information). For both RW-AFM and 3Q states those with effective $\rightleftarrows$ interlayer coupling (Fig.~\ref{fig:RWAFM_shift}a,c) exhibit a lower total energy compared to the corresponding states with $\rightrightarrows$ coupling between the layers (Fig.~\ref{fig:RWAFM_shift}b,d). 

In contrast to the experimental result, where the RW-AFM state is present in the domains and the 3Q state is only realized in the junctions, we find that the 3Q state (Fig.~~\ref{fig:RWAFM_shift}a) has the lowest DFT total energy, i.e.~it is by 15~meV/Mn atom lower than the RW-AFM state \footnote{Note, that we have checked that spin spiral states, which comprise the fundamental solution of the Heisenberg model on a periodic lattice, are higher in energy than the lowest RW-AFM and 3Q state \cite{GutzeitPhD}.}. However, we have not yet taken into account that the RW-AFM state breaks the three-fold symmetry of the substrate and that the Mn atoms may shift out of the hollow-site adsorption sites~\cite{ZahnerAX2024}. 

Upon allowing in our DFT calculations the lateral coordinates of the Mn atoms to relax within the RW-AFM state, there is a large shift of the top Mn layer and a significant energy gain of about 57~meV/Mn atom (see Fig.~\ref{fig:RWAFM_shift}e). As a result this shifted RW-AFM$\rightleftarrows$ state is about 42~meV/Mn atom lower than the corresponding 3Q$\rightleftarrows$ state (Fig.~\ref{fig:RWAFM_shift}a). The Mn atoms of the upper magnetic layer shift by about 1 {\AA} along the [11$\overline{2}$]-direction (see Fig.~\ref{fig:RWAFM_shift}c vs.~Fig.~\ref{fig:RWAFM_shift}e) thereby getting closer to the atoms with opposite spin in the subsurface Mn layer and hence leading to a strengthening of the antiferromagnetic bonds (see also Supplementary Information). 
Note, that the RW-AFM$\rightrightarrows$ structure with its net FM coupling of the moments between the two layers also experiences a lateral shift (Fig.~\ref{fig:RWAFM_shift}f). However, the effect is much less pronounced than for the RW-AFM$\rightleftarrows$ state with the Mn atoms of the top layer being displaced by only 0.26~{\AA} and a much smaller energy gain of about 17~meV/Mn atom with respect to the unshifted state (Fig.~\ref{fig:RWAFM_shift}d).
In conclusion our DFT calculations show that the RW-AFM$\rightleftarrows$ state with the shifted top layer is the lowest energy state, in agreement with the experimental findings.

\begin{figure}[htb]
    \centering
    \includegraphics[width=1.0\columnwidth]{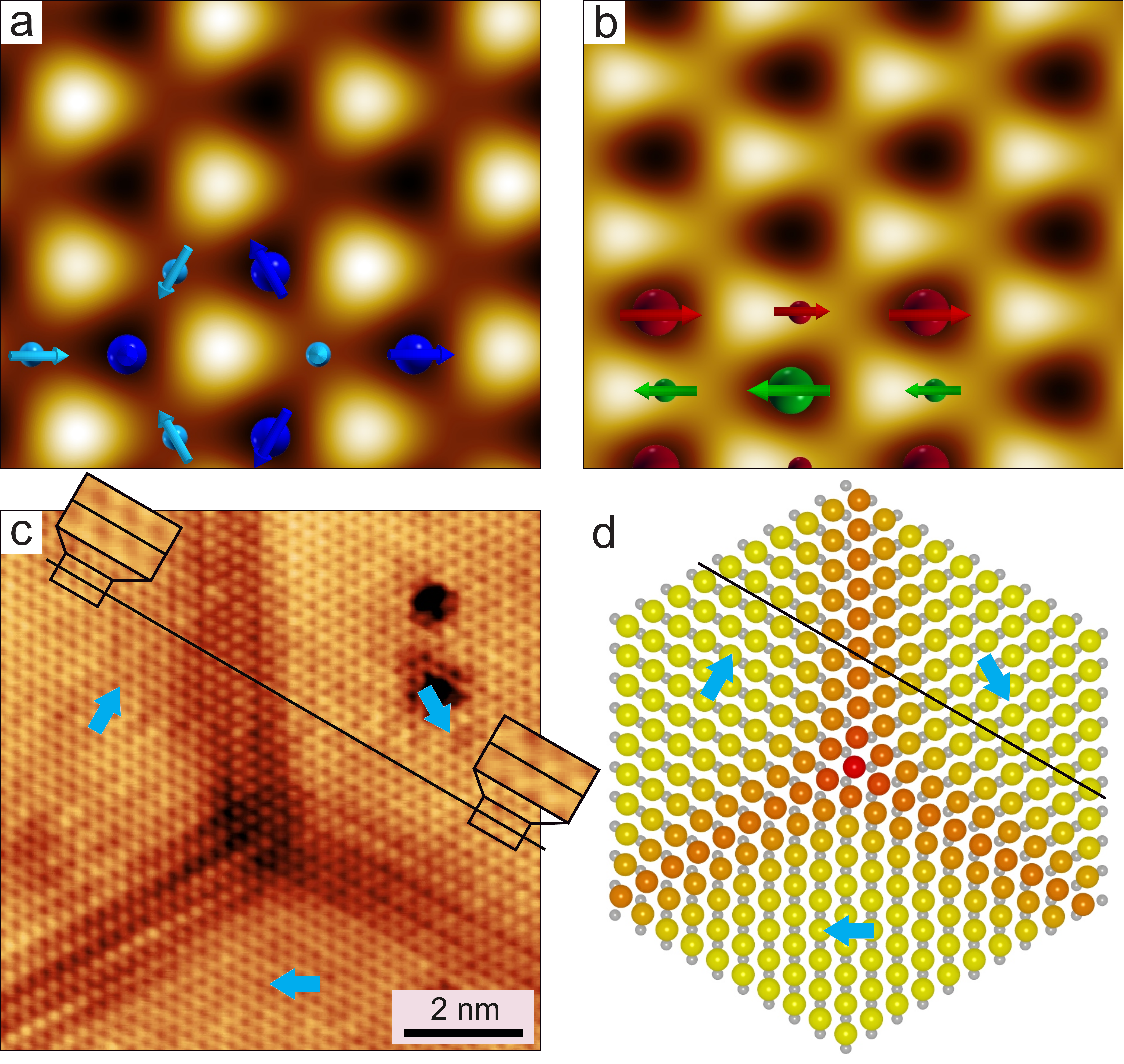}
    \caption{\textbf{Magnetism-driven structural shift.} 
    \textbf{a,}~STM image calculated via DFT for the 3Q$\rightleftarrows$ state (Fig.~\ref{fig:RWAFM_shift}a). Large (small) spheres indicate atoms of the top (bottom) Mn layer. 
    \textbf{b,}~STM image calculated via DFT for the shifted RW-AFM$\rightleftarrows$ state (Fig.~\ref{fig:RWAFM_shift}e).  
    \textbf{c,}~Atomic resolution constant-current STM image of a Y$^*$-junction; the black line is placed across an atomic row (dark spots) in the right domain, and extrapolation to the top domain confirms the different relative shifts for the top Mn layer, as indicated by cyan arrows ($U = +10$~mV, $I = 5$~nA, $T = 4$~K).
    \textbf{d,}~Atomistic model of a Y$^*$-junction where the atom positions vary from hollow-site in the center (3Q) to bridge site in the RW-AFM domains; the color code illustrates the size of the shift,
    which leads to a structural handedness with compressive strain on one side and tensile strain on the other side of each domain wall.
    } 
	\label{fig:atomic}
\end{figure}

For the two magnetic states of lowest energy, i.e.\ the 3Q$\rightleftarrows$ state and the shifted RW-AFM$\rightleftarrows$ state (Figs.~\ref{fig:RWAFM_shift}a,e), we have calculated spin-averaged STM images based on the Tersoff-Hamann model~\cite{Tersoff1985} (see methods). Figures~\ref{fig:atomic}a,b show that in both cases we find anticorrugation, i.e.\ at the position of the surface atoms the signal is minimal~\cite{Heinze1998}. Whereas the image of the 3Q state in Fig.~\ref{fig:atomic}a shows threefold symmetry, the image of the RW-AFM in Fig.~\ref{fig:atomic}b reflects the two-fold symmetry of the uniaxial magnetic state, even in the displayed spin-averaged case. Figure~\ref{fig:atomic}c shows an  atomically resolved constant-current STM image of a Y$^*$-junction and we find that also here the pattern changes between the center of the junction and the extended RW-AFM domains.

Assuming that the positions of lowest signal indicate the atom positions, as seen in the DFT calculations, we find by extrapolation of the atomic rows in the RW-AFM domains that the top-layer Mn atoms are not aligned across adjacent orientational domains, providing experimental evidence for the calculated lateral shift of the top layer, see cyan arrows. A model of the shifted top Mn layer around a Y$^*$ junction is displayed in Fig.~\ref{fig:atomic}d, where the atoms of the three orientational domains are shifted from the hollow sites to bridge sites, whereas atoms in the center of the junction remain in perfect hollow site positions.

\section*{Characterization and Handedness of the Junctions}

\begin{figure}[htb]
    \centering
	\includegraphics[width=0.95\columnwidth]{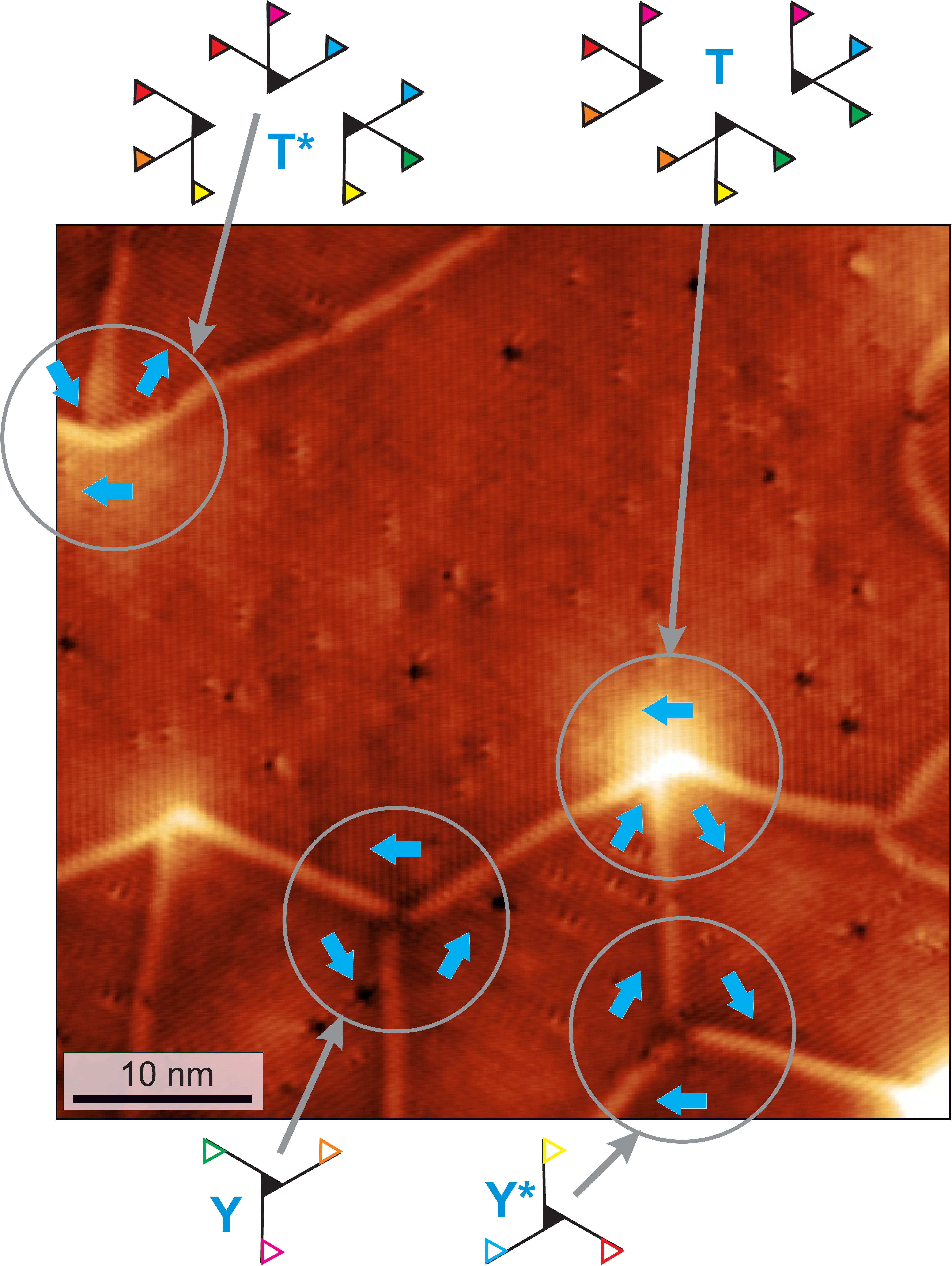}
    \caption{\textbf{Chirality of the triple-junctions.}
    SP-STM constant-current image of several triple-junctions, the different types are labelled; cyan arrows indicate the shift direction of the top Mn layer for the different orientational domains ($\Delta z=50$~pm). Drawing lines along the $120^\circ$ domain walls in different junctions results in an asymmetric intersection (see sketches) of the three domain walls due to the structural shift. Sketches show all possible triple-junctions that can occur with $120^\circ$ domain walls, the T/T$^*$-junctions occur in three orientations. To form a network, a pair of empty (Y/Y$^*$-junction) and filled same color triangles (T/T$^*$-junction) need to be connected, i.e., any straight wall must have a Y/Y$^*$-junction at one end and a T/T$^*$-junction at the other end respectively which is the only possible way of forming the network.
    (Measurement parameters: $U = +10$\,mV, $I = 5$\,nA, $T=4$\,K). 
    } 
	\label{fig:handedness}
\end{figure}

Close inspection of the Y and T-junctions shows that the shift of the top Mn layer is also reflected by the apparent position of the domain walls, as evident at the triple-junctions present in the SP-STM image of Fig.~\ref{fig:handedness}. Two mirror-symmetric versions of each junction type are indicated, i.e.\ Y and Y$^*$ and T and T$^*$. When the positions of the straight $120^{\circ}$ walls are extrapolated to the junction's center one can see that they do not merge in one point but instead span a triangle, see sketches for the individual junction types. This demonstrates a handedness of the junction and we find the same chirality for all Y-junctions, and opposite chirality for all Y$^*$-junctions. The same is true for T and T$^*$-junctions, which can occur in three orientations each.

This chirality is a consequence of the lateral shift of the top layer (see cyan arrows for the respective shift directions), which induces significant strain at the position of the domain walls. Also the path of the domain wall is strongly connected to the orientation of the adjacent domains, which has been attributed to the magneto-elastic energy in a theoretical study of a system with orthogonal orientational domains~\cite{ConsoloRM2023}. 
Due to the unique domain configurations around Y$^{(*)}$- and T$^{(*)}$-junctions these triple-junctions always come in Y$^{(*)}$-T$^{(*)}$pairs, and the possible pairing configurations within the network are indicated by the color of the triangles.

\begin{figure}[htb]
    \centering
	\includegraphics[width=0.95\columnwidth]{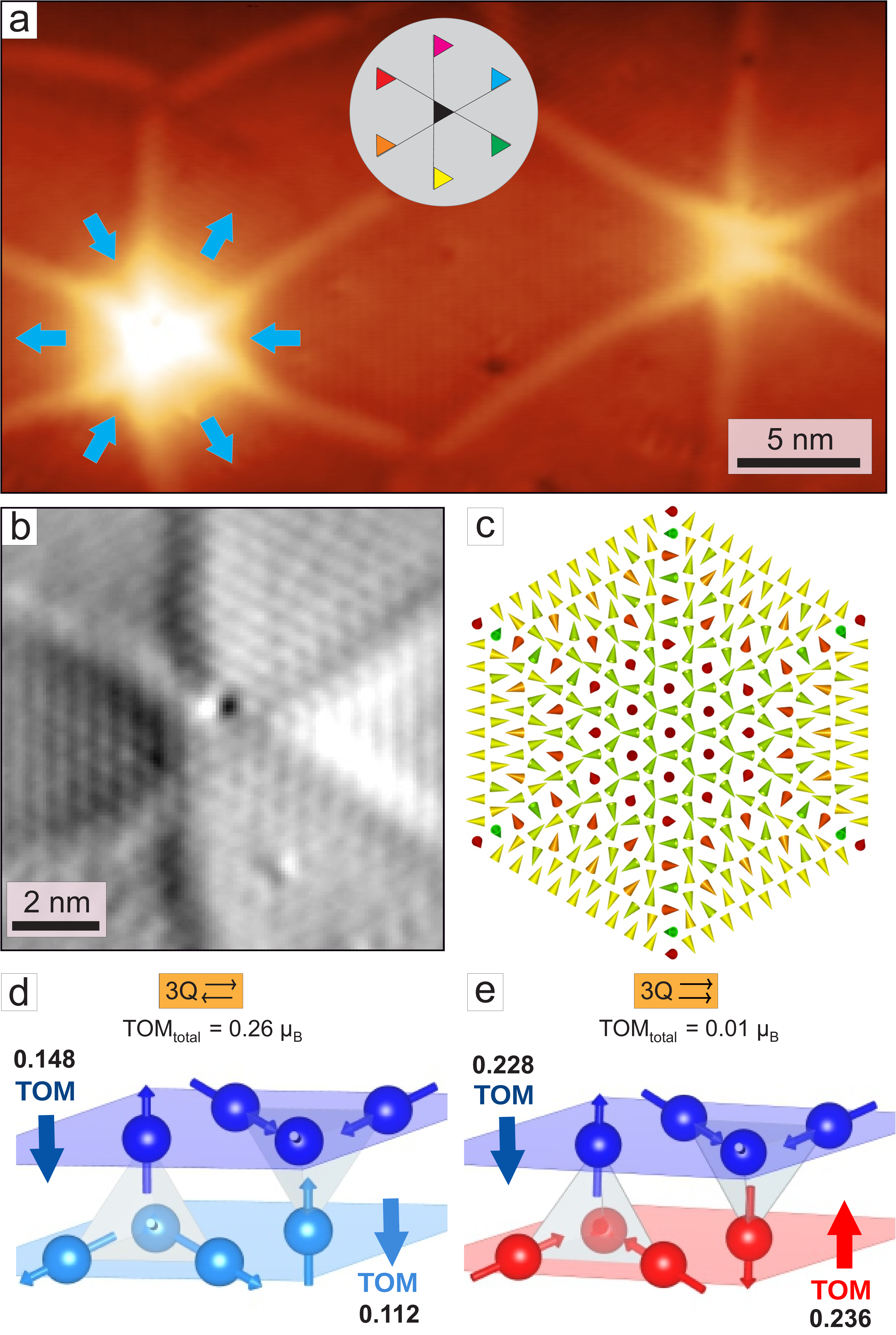}
    \caption{\textbf{Hexa-junctions and topological orbital moments.} 
    \textbf{a},~SP-STM constant-current image of the antiferromagnetic domain wall network with two hexa-junctions located at Ar bubbles ($\Delta z=150$~pm), cyan arrows indicate the shift direction of the top Mn layer. 
    \textbf{b},~Simultaneously measured current map of the left hexa-junction of (a), showing the hexagonal magnetic pattern in the junction center indicative of the 3Q state ($\Delta I=\pm 200$\,pA). (Measurement parameters: $U = +10$\,mV, $I = 1$\,nA, $T= 4$\,K).
    \textbf{c}, Spin model of a hexa-junction.
    \textbf{d,e}, Sketches illustrating the two possible 3Q configurations for a double-layer (cf.~Fig.~\ref{fig:RWAFM_shift}a,b); the arrows indicate the DFT calculated direction of the TOM for each individual layer; values are the DFT calculated layer-resolved and total TOMs for each configuration, and are given in $\mu_B$/magnetic unit cell.
    }
	\label{fig:hexajunctions}
\end{figure}

After the  analysis of the magnetic state in the triple-junctions we turn to the question why so many domain walls appear in the first place. While in ferromagnets the incorporation of domain walls can lower the total energy due to a reduction of the stray field contributions, in antiferromagnets this is not the case and domain walls in ultra-thin films always have a positive energy contribution. We find that in our sample the Ar bubbles connect up to six walls, see Fig.\ref{fig:network} and Fig.~\ref{fig:hexajunctions}, and conclude that they initiate the formation of the observed domain wall network. Indeed, samples without Ar bubbles do not show such a high domain wall density (see Suppl. Fig.~1). 

In our double-layer system the RW-AFM state leads to a shift of the atoms. A shift of atoms is also expected to happen locally in the vicinity of Ar bubbles due to the strain~\cite{GsellS1998}, even above the magnetic transition temperature. While the induced local curvature of the Mn double-layer leaves the tangential atom-atom distance nearly unaffected, the radial distances slightly elongate. This locally breaks the threefold symmetry at the slopes of the Ar bubble. We speculate that upon cooling through the critical magnetic temperature this local strain field
triggers the formation of one specific orientational domain on the sides of the Ar bubble, similar to what has been reported before for a preferential orientation of a uniaxial adsorbate arrangement around an Argon bubble~\cite{GsellS1998,JakobJCP2001}. Indeed, all Ar bubbles have those RW-AFM domains around them, which show a radial shift of the top Mn layer, see cyan arrows in Fig.~\ref{fig:hexajunctions}a. The experimentally observed asymmetry of the wall positions with respect to the center of the junction reflects the strain due to this shift of the top Mn layer.

Due to the comparably large height of the Ar bubble the magnetic pattern of the hexa-junction can better be seen in the simultaneously acquired current map, as shown in Fig.~\ref{fig:hexajunctions}b for the left hexa-junction. Again, the pattern in the junction center is hexagonal, indicative of a local 3Q state. Such a configuration with six domain walls is also illustrated in the spin model of Fig.~\ref{fig:hexajunctions}c (see methods). The majority of the experimentally found hexa-junctions are better characterized as pairs of T-T$^*$-junctions (see for instance the right hexa-junction in Fig.~\ref{fig:hexajunctions}a) (see also Suppl. Fig.~3).

\section*{Emergent Topological Properties of the Junctions}

The triple- and the hexa-junctions, which form at meeting points of the orientational RW-AFM domains, locally exhibit the 3Q state. This intriguing three-dimensional spin structure has a vanishing net spin moment. However, it is predicted to exhibit 
topological orbital moments
(TOMs)~\cite{Hanke2016,Grytsiuk2020,Nickel2023}, which are susceptible to applied magnetic fields and lead to the spontaneous topological Hall effect~\cite{TakagiNN2023,ParkNC2023}. TOMs occur due to electron motion in the emergent magnetic field which is proportional to the finite scalar spin chirality of a non-coplanar spin structure such as the 3Q state~\cite{Hanke2016,Haldar2021}. Two inverted 3Q states with opposite directions of TOMs along the surface normal exist, often referred to as ‘all-in’ and ‘all-out’ 3Q state~\cite{TakagiNN2023,ParkNC2023,Nickel2023}. 

For systems beyond a single layer, the 3Q state can occur both with an effective ferromagnetic ($\rightrightarrows$) or antiferromagnetic ($\rightleftarrows$) coupling between the layers (Fig.~\ref{fig:hexajunctions}d,e). The absolute value of the TOM per Mn layer is larger for the 3Q$\rightrightarrows$ state, however, the TOMs in adjacent layers point in opposite directions and the total TOM is almost zero (Fig.~\ref{fig:hexajunctions}e). This is a consequence of combining an 'all-in' with an 'all-out' configuration in the two layers.

In contrast, for the energetically more favorable 3Q$\rightleftarrows$ state (Fig.~\ref{fig:hexajunctions}d) the TOMs of the two layers point in the same direction, and add up to a total TOM of $0.26~\mu_B$/magnetic unit cell. In this configuration, the 3Q state is realized not only within each Mn layer but also between the layers (see tetrahedron in Fig.~6d). Whereas the TOM vanishes for the collinear RW-AFM domains and the coplanar 2Q superposition domain walls, all the triple- and hexa-junctions exhibit the non-coplanar 3Q state and therefore also an associated local TOM.

\section*{Discussion}

Our combined experimental and theoretical work shows that the formation of a domain wall network in the Mn double-layer on Ir(111) can be induced by the incorporation of local strain. Due to the presence of three symmetry-equivalent orientational magnetic domains, this network forms triple- and hexa-junctions, which locally exhibit non-coplanar spin textures. The use of SP-STM has enabled us to correlate the local magnetic properties to the strain fields induced by the localized Ar bubbles as well as the antiferromagnetic domain walls themselves. DFT calculations have revealed the interplay of magnetic energies with the shift of the top layer and the size of the TOMs arising for the non-collinear magnetic state at the antiferromagnetic domain wall junctions. 

We anticipate that beyond the fundamental interest in antiferromagnetic domain walls and junctions such networks are interesting candidates for novel computing schemes as they provide several different unique transport properties: (i) the uniaxial collinear magnetic domains are expected to show transport characteristics depending on the angle of the lateral current with respect to the Néel vector, (ii) likely the non-collinear magnetic state in the $120^{\circ}$ superposition domain walls strongly interacts with current, (iii) the non-coplanar 3Q junctions give rise to the topological Hall effect. While the latter is expected to cancel for zero-field cooled samples, we anticipate that the effect can be maximized by field-cooling, which initiates the 3Q state with a TOM parallel to the applied magnetic field. However, while strong interaction with currents are likely, the network itself is expected to be strongly pinned due to the strain-induced hexa-junctions.
We expect that such domain wall networks can be created by strain in various antiferromagnetic materials, opening the possibility to study their response to lateral currents in view of future spintronics applications.

\clearpage

\bibliography{Mn_DL_Ir111_arxiv}

\clearpage

\section*{Methods}

\textbf{Sample Preparation.}
Samples were prepared in ultra-high vacuum and transferred \emph{in-situ} to a low-temperature STM. Ir(111) crystals were cleaned by cycles of Ar-ion etching ($1.5$\,keV) and subsequent annealing ($T \approx 1600$\,K). To obtain Ar bubbles below the surface the last annealing step was modified, and a lower annealing temperature of about $1300$\,K was used~\cite{GsellS1998}. Mn was evaporated from a Knudsen cell held at $T \approx 1050$\,K, leading to a deposition rate of about $0.15$~atomic layers per minute. The Ir(111) substrate was kept at around 470K during Mn deposition.

\textbf{(Spin-Polarized) Scanning Tunneling Microscopy.}
To obtain information on the magnetic properties of a surface we employ SP-STM, which exploits the spin-polarization of the tunnel current between two magnetic electrodes. The spin-polarized tunnel current scales with the cosine of the angle between tip and sample magnetization directions~\cite{Wortmann2001,WiesendangerRMP2009,BergmannJPCM2014}. Here we have used a Cr-bulk tip (except for Suppl. Fig.~3c, for which an Fe-coated W tip was used). 
Throughout this work we refer to measurements with and without spin-polarized
contrast as SP-STM images and STM images, respectively.

The RW-AFM state always appears as stripes, and the magnetic corrugation amplitude depends on the relative tip magnetization direction: when the tip magnetization direction is along the quantization axis of the RW-AFM state we obtain maximal magnetic signal, whereas an orthogonal orientation results in a vanishing magnetic contrast.
The 3Q state is a non-coplanar state and regardless of the tip magnetization direction this leads to spatial variation of the spin-polarized contribution to the tunnel current. The possible patterns range from a hexagonal p(2x2) pattern to a stripe p(2x1) pattern~\cite{Spethmann2020}. 

The fact, that the domain walls can be seen very prominently in the STM measurements is dominantly due to the spin-averaged contribution to the tunnel current. Different magnetoresistance effects come to mind, for instance the non-collinear magnetoresistance effect, where a signal difference is observed for collinear versus non-collinear magnetic order~\cite{Hanneken2015}. However, for this system likely the dominating effect is the local strain at the domain walls. Because the Mn atoms in the RW-AFM domains are located roughly in the bridge sites, and those at the junctions or the domain walls are more closer to three-fold hollow sites, this difference in adsorption site is expected to be the main origin for the strong signal difference.

Maps of differential conductance (d$I$/d$U$) were obtained simultaneously to the constant-current images with a lock-in amplifier, and the sample bias voltage was modulated with a peak to peak amplitude of about $10\%$ at a frequency on the order of $3-5$~kHz.

The presented STM images are (plane-fitted) raw data (except when specified otherwise) and were obtained at zero magnetic field (note that Fig.~1a,b, Fig.~2c,d, Fig.~5, Fig.~6a,b are in the remanent state; however, we do not see any difference between remanent and virgin state).

\textbf{Spin Models.}
To illustrate the observed spin structures we have employed spin dynamics simulations. In the Heisenberg model the RW-AFM state and the 3Q state are degenerate, both in a monolayer and a double-layer. An energy difference can arise due to higher-order interactions~\cite{Kurz2001,Spethmann2020}. Because an extraction of higher-order interactions from DFT calculations for double-layers has not been possible up to now, the relevant parameters that would serve as input parameters to realistically describe our system are missing. For this reason we have used a monolayer setup with generic parameters to illustrate the different building blocks for the network, without the aspiration to perfectly describe the system under study.
To set up the system we have used the following Hamiltonian~\cite{SpethmannNC2021}

\begin{equation}
\begin{split}
     H  &=  - \sum_{i,j} J_{ij} (\mathbf{S}_{i} \cdot \mathbf{S}_{j} )
       - \sum\limits_{i,j} B_{ij} ( \mathbf{s}_i \cdot \mathbf{s}_j)^2 \\
       & - \sum_i K_u (\mathbf{s}_i \cdot \hat{\mathbf{z}})^2
       - J_{\rm ASE} \sum_{ij} (\mathbf{s}_i \cdot \mathbf{d}_{ij}) (\mathbf{s}_j \cdot \mathbf{d}_{ij})
\end{split}
\end{equation}
          
including nearest neighbor isotropic exchange $J_1$, next-nearest neighbor isotropic exchange $J_2$, ..., the biquadratic (or 2-site-4-spin) higher-order term $B$, an unaxial magnetocrystalline anisotropy $K_u$, and the anisotropic symmetric exchange $J_{\rm ASE}$, where $\mathbf{S}_{i}$ denotes a spin at a lattice site specified by i, $\hat{\mathbf{z}}$ is the unit vector perpendicular to the surface and $\mathbf{d}_{ij}$ is the normalized connection vector between the lattice sites i and j. The spin dynamics simulations presented in the manuscript were done with Monte Crystal 3.2.0, which can be found on github~\cite{MonteCrystal}.

The parameters for the different images vary slightly and are: $J_1 = -25$, $J_2 =-5$, $J_{\rm ASE} = +0.1$, $B = +0.25 \pm 0.15$, $K_u = -0.4$ (all in meV/atom); Fig.~\ref{fig:network}d is a cut-out of a much larger simulation box and in the case of Fig.~\ref{fig:walls}e and Fig.~\ref{fig:hexajunctions}c the spins of some edges were fixed throughout the simulation; Fig.~\ref{fig:atomic}d is obtained from the spin model in Fig.~\ref{fig:walls}d (rotated by 180$^\circ$): atoms within the 1Q (RW-AFM) domain atoms are positioned in bridge sites, atoms in the 3Q junction center are positioned in hollow sites, in-between atoms are shifted according to their 1Q components; linear color scale red-yellow by atom shift distance.

\textbf{Density functional theory calculations.}
We employed density functional theory (DFT) using the projected augmented wave (PAW) method~\cite{paw} as implemented in the {\tt VASP} code~\cite{vasp,Kresse1996,Kresse1999} to perform both geometry optimizations and calculations of total energies for different magnetic states of Mn double-layers (DLs) on Ir(111). In all calculations the theoretical in-plane lattice constant of Ir within the generalized gradient approximation (GGA) was used which amounts to 2.75 {\AA}~\cite{vonBergmann2006} and a large energy cutoff of 400 eV was chosen. Exchange-correlation effects were included by means of the GGA potential with the interpolation developed by Perdew, Burke and Ernzerhof (PBE)~\cite{PBE}. For further computational details of these calculations see the Supplementary Information.

In order to calculate STM 
images based on DFT (see Fig.~\ref{fig:atomic}a,b) we applied the Tersoff-Hamann model~\cite{Tersoff1985}.
These DFT calculations were performed for the 
laterally relaxed 
RW-AFM$\rightleftarrows$
state as well as for the 3Q$\rightleftarrows$ state of the Mn double-layer on Ir(111) using the full-potential linearized augmented planewave (FLAPW) method~\cite{Wimmer1981} as implemented in the {\tt FLEUR} code~\cite{FLEUR}. Exchange-correlation effects were included in the local density approximation using the parameterization of Vosko, Wilk and Nusair~\cite{vwn}. Besides applying the two-atomic rectangular (four-atomic hexagonal) unit cell of the RW-AFM ($3Q$) state, an asymmetric slab with nine Ir layers and a Mn DL and 247 $k$-points including the $\overline{\Gamma}$-point (288 $k$-points) in the irreducible part of the BZ were used. The muffin tin radii were set to 
2.16 a.u. for Mn in the 
shifted RW-AFM$\rightleftarrows$ state and to 2.31 a.u. for Ir. Moreover, a large cutoff of $k_{\text{max}}$=4.1 a.u.$^{-1}$ was chosen to ensure convergence with respect to the basis functions. The STM images were calculated for an energy window corresponding to a
bias voltage of
$+100$~meV 
and at a height of 3 {\AA} above the surface. 

We have further employed the FLAPW based
{\tt FLEUR}
code to check
the results for the shifted RW-AFM$\rightleftarrows$ state obtained via {\tt VASP}. Applying the above mentioned geometric {\tt FLEUR} setup, i.e. a two-atomic rectangular unit cell, and the PBE~\cite{PBE} exchange correlation potential, the energy gain of the laterally relaxed structure with respect to the unshifted state is calculated as $-64.18$ meV/Mn atom, whereas the corresponding {\tt VASP} value amounts to $-57.36$ meV/Mn atom (cf.~Fig.\ref{fig:RWAFM_shift}e). Hence, the energy gain obtained via {\tt FLEUR} is on a similar order of magnitude.



\section*{Acknowledgements}
This project has received funding from the European Union’s Horizon 2020 research and innovation programme under the Marie Skłodowska-Curie grant agreement No 955671.
K.v.B.\ gratefully acknowledges financial support from the Deutsche Forschungsgemeinschaft (DFG, German Research Foundation) via projects no. 402843438 and no. 418425860. 
M.G., So.H., and St.H. gratefully acknowledge financial support from the Deutsche Forschungsgemeinschaft (DFG, German Research Foundation) via SPP2137 “Skyrmionics” (project no.~462602351) and project no.~418425860 and supercomputing
resources provided by 
the North-German Supercomputing Alliance (HLRN).

\section*{Author contributions}
V.S.\ and F.Z.\ prepared the samples. V.S.\ performed the measurements. V.S., A.R.-S., and K.v.B.\ analyzed the data. A.K.\ performed the spin dynamics simulations and made the spin models. M.G and So.H. performed the DFT calculations. 
M.G., So.H., and St.H. analyzed the DFT data.
K.v.B., M.G., and St.H. wrote the manuscript. V.S., M.G., A.R.-S., So.H., F.Z., R.W., A.K., St.H., and K.v.B.\ discussed the results and commented on the manuscript.



\clearpage

\setcounter{figure}{0}
\renewcommand{\thefigure}{Supplementary \arabic{figure}}

\onecolumngrid

\section*{Supplementary Figures}

\begin{figure}[htb]
	\centering
    \includegraphics[width=0.85\textwidth]{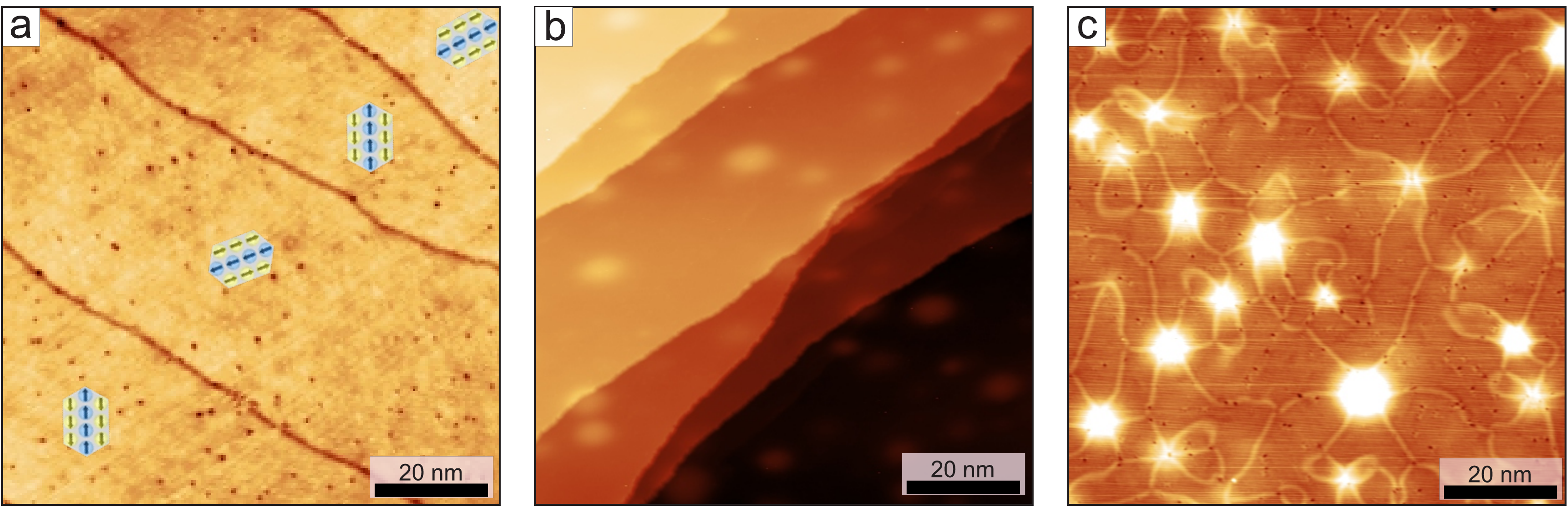}
	\caption{\textbf{Impact of Ar bubbles on domain wall density.} 
    \textbf{a},~Constant-current STM image of the Mn double-layer on a broad Ir(111) terrace ($\Delta z=25$~pm). In contrast to the other samples with Ar bubbles, in this preparation the Ir(111) substrate was annealed at a higher temperature of $T \approx 1600$\,K before the deposition of Mn. Dark lines indicate domain walls between orientational domains of the RW-AFM state; note that due to a different sample bias voltage the domain walls appear dark in this measurement. (Measurement parameters: $U = +500$\,mV, $I = 1$\,nA; $T=10$\,K). 
    \textbf{b},~Constant-current STM image of Ar bubbles in the uncovered Ir(111) surface ($\Delta z=1.2$~nm). After the final Ar-ion etching step the Ir crystal was annealed at a reduced temperature of $T \approx 1300$\,K twice for 40 seconds each, i.e., the same procedure as for the sample preparations shown in the main text figures, only without subsequent deposition of Mn. (Measurement parameters: $U = +250$\,mV, $I = 0.5$\,nA; $T=300$,K).
    \textbf{c},~Constant-current STM image of the Mn double-layer on an Ir(111) surface with Ar bubbles ($\Delta z=144$~pm); bright lines indicate domain walls between orientational domains of the RW-AFM state, which form a network induced by the strain around the Ar bubbles. (Measurement parameters: $U = +10$\,mV, $I = 1$\,nA; $T=4$\,K).
    }
	\label{fig:DW_density}
\end{figure}


\begin{figure*}[htb]
	\centering
    \includegraphics[width=0.6\textwidth]{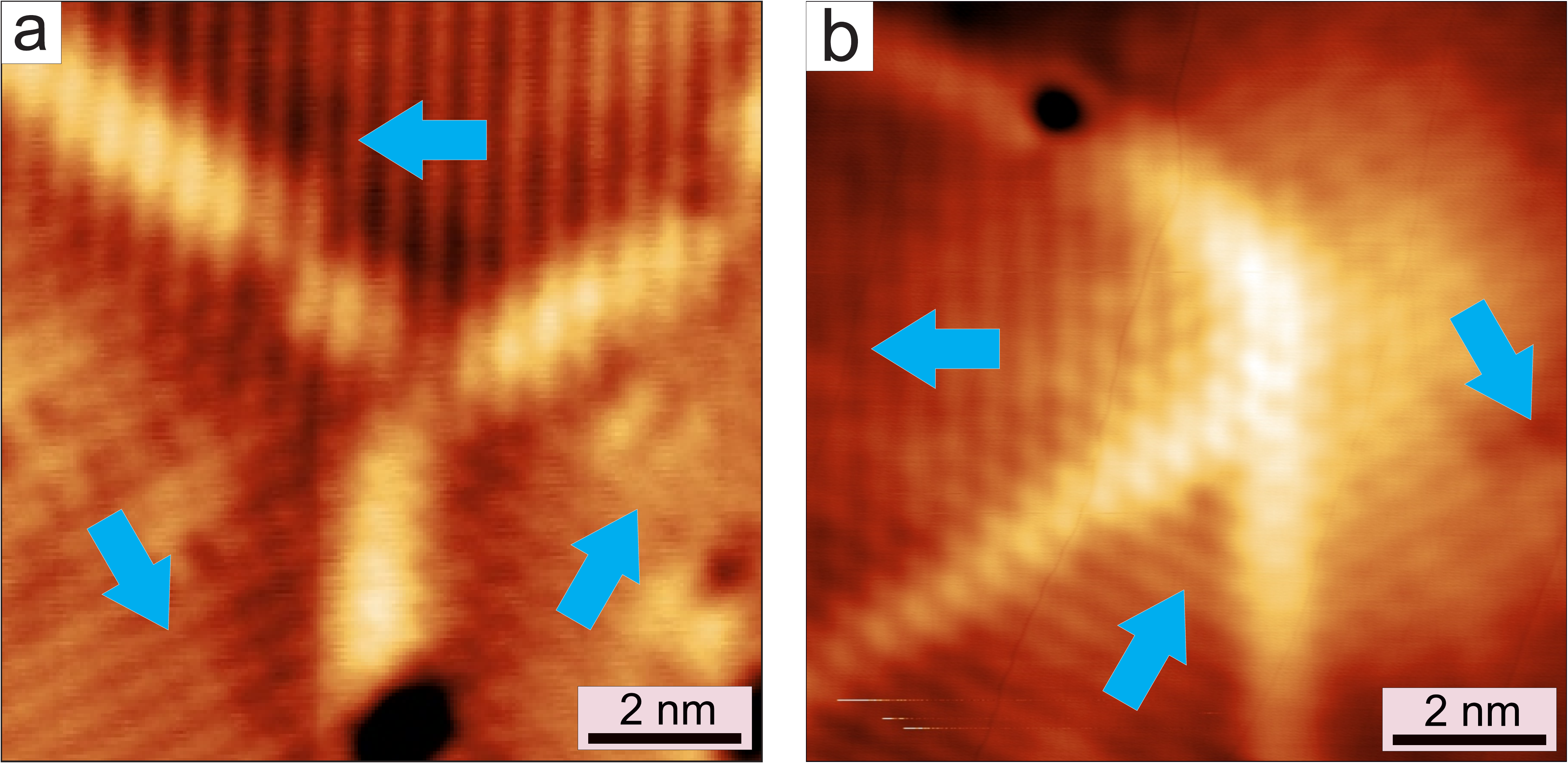}
	\caption{\textbf{Additional data demonstrating the occurrence of the 3Q state at the triple-junctions.} \textbf{a},\textbf{b},~Constant-current SP-STM of a Y- and a T$^*$-junction, respectively (a: $\Delta z=21$\,pm, b: $\Delta z=44$\,pm), exhibiting the hexagonal magnetic pattern of the 3Q state in their centers. The cyan arrows indicate the structural shift direction for the different orientational domains. (Measurement parameters: a: $U = +10$\,mV, $I = 1$\,nA, $B= 0$\,T; b: $U = +10$\,mV, $I = 7$\,nA, $B= +2.5$\,T; both: Fe-coated W-tip, $T=8$\,K.)
    }
	\label{fig:Y_and_T}
\end{figure*}


\begin{figure}[htb]
	\centering
	\includegraphics[width=0.9\textwidth]{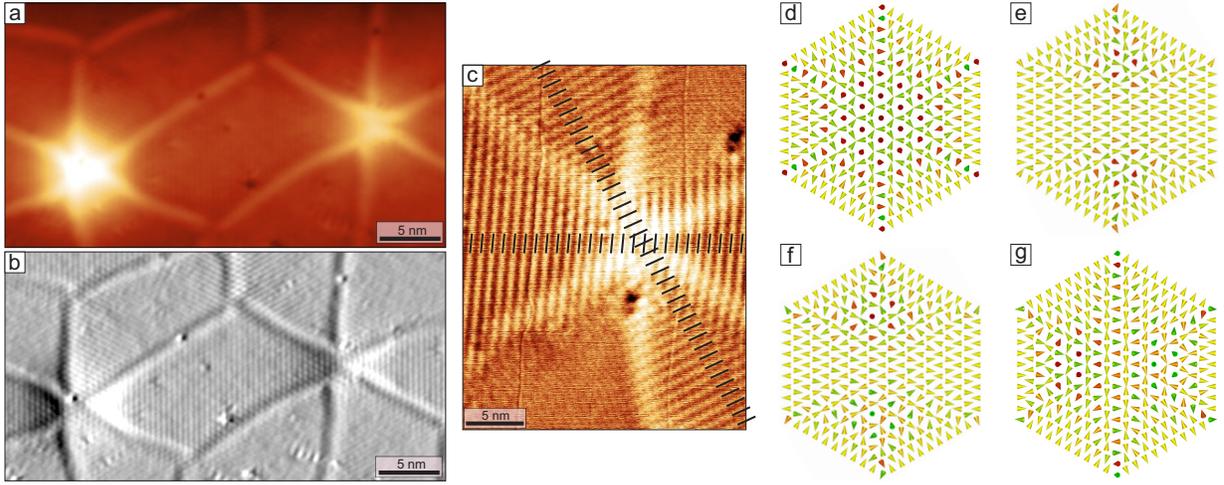}
	\caption{\textbf{Hexa-junction spin texture at the Ar bubbles.} 
    \textbf{a},~SP-STM constant-current image of two hexa-junctions in the domain wall network (same data as displayed in Fig.~\ref{fig:hexajunctions}a, slightly larger field of view, $\Delta z = 150$\,pm). 
    \textbf{b},~Current map obtained simultaneously to the SP-STM image of (a), in which the hexagonal magnetic pattern in the center of the hexa-junctions can be seen ($\Delta I = \pm150$\,pA). (Measurement parameters: $U = +10$\,mV, $I = 1$\,nA, $T=4$\,K, Cr-bulk tip). 
    \textbf{c},~SP-STM constant-current image of a different hexa-junction (same one as displayed in Fig.~6b), after a highpass filter with a cut-off frequency corresponding to $0.55$\,nm was applied to the raw data in order to remove the height variation of the Ar bubble and enhance the visibility of the magnetic state (now $\Delta z=4$~pm). Equally spaced black lines indicate the maxima of magnetic contrast of the RW-AFM and demonstrate that two pairs of opposite RW-AFM domains across the Ar bubble are in-phase; no magnetic contrast was obtained for the third orientational domain. (Measurement parameters: $U = +10$\,mV, $I = 1$\,nA, $T=8$\,K, Fe-coated W tip.) 
    \textbf{d},~Spin model with six RW-AFM domains, with the constraint that three pairs of opposite RW-AFM domains are in phase, resulting in a perfect 3Q state in the center of the hexa-junction; the hexagonal magnetic pattern of the hexa-junction in (c) and the spatial distribution thereof are in good agreement with this spin model. 
    \textbf{e},~Spin model with six RW-AFM domains, with the constraint that three pairs of opposite RW-AFM domains are in phase, resulting in a different spin texture in the center of the hexa-junction, which can be better described as a pair of T-/T$^*$-triple-junctions; a T-/T$^*$-triple-junctions is present at the right side of the measurement in (a), but it has not been possible to identify the spin texture at its center. 
    \textbf{f},\textbf{g},~Spin models with six RW-AFM domains, with the constraint that some pairs of opposite RW-AFM domains are out of phase, resulting in pairs of T-/T$^*$-triple-junctions with complex spin textures in their centers. 
    } 
	\label{fig:phase_check}
\end{figure}

\end{document}